\documentclass{svproc}
\usepackage{url}

\usepackage[colorlinks=true,citecolor=blue,linkcolor=blue,breaklinks=true]{hyperref}
\usepackage{amsmath,amssymb}
\usepackage{epsfig}  
\usepackage{graphicx}                
\usepackage{url}
\usepackage{color}
\usepackage{multirow}
\usepackage{placeins}
\usepackage[dvipsnames]{xcolor}

\definecolor{red}{rgb}{1.0, 0, 0}
\definecolor{jblue}  {RGB}{20,50,100}
\definecolor{npurple}  {RGB} {153, 51, 204}  
\definecolor{wred}   {RGB}{217,0,56}
\definecolor{white}   {RGB}{255,255,255}

\usepackage{cleveref}
\usepackage{subfigure}
\usepackage{lipsum}
\usepackage{gensymb}
\renewcommand{\vec}[1]{{\mathbf{#1}}}

\newcommand{\w}{\omega}
\newcommand{\lam}{\lambda}

\begin{document}
\mainmatter              % start of a contribution
\title{Supernova neutrinos: fast flavor conversions near the core}
\titlerunning{Fast Flavor Conversions}  % abbreviated title (for running head)
\author{Manibrata Sen }
%
%\authorrunning{Manibrata Sen} % abbreviated author list (for running head)
%Craig Chambers, Kim B. Bruce, and Elisa Bertino}
%
\institute{Tata Institute of Fundamental Research,
             Homi Bhabha Road, Mumbai, 400005, India.\\
\email{manibrata.sen@gmail.com }.
%\texttt{http://users/\homedir iekeland/web/welcome.html}
}

\maketitle              % typeset the title of the contribution

\begin{abstract}
Neutrino flux streaming from a supernova 
can undergo rapid flavor conversions almost immediately above the 
core. Focusing on this region, we study these 
\emph{fast conversions} using a linear stability analysis.
We find that, for realistic angular distributions of neutrinos, fast conversions can 
occur within a few nanoseconds in regions just above the neutrinosphere. Our results also show that neutrinos travelling towards the core make fast conversions 
more rapid. These conversions, if they exist, can have significant implications for supernova explosion mechanism and nucleosynthesis.
\end{abstract}
\begin{flushright}
 \scriptsize{TIFR/TH/17-08}
\end{flushright}

\section{Introduction}
\label{sec:1} 
 Core collapse supernovae (SN) offer a fascinating environment to study neutrino flavor
 evolution in dense environments.  Very recently, it was predicted
 that fast flavor conversions, occurring with a rate $\mu\sim\sqrt{2}G_Fn_\nu$ ($n_\nu$ is the neutrino density), can 
 happen very near the SN core, as opposed to the well-known collective effects, occurring at $r\sim\mathcal{O}(10^2)$km
 \cite{Duan:2006an,Hannestad:2006nj} or the MSW effect at $r~\sim\mathcal{O}(10^3)$km \cite{Wolfenstein:1977ue,Mikheev:1986gs}. A necessary condition for this seemed to be a non-trivial
 angular distribution in the neutrino emission spectrum \cite{Sawyer:2015dsa,Chakraborty:2016lct}. Based on these claims, we make a 
 detailed study of these fast flavor conversions. We focus on regions \emph{close}
 to the SN core and hence model the source as a flat geometry as shown in the left panel of Fig.\,\ref{fig:2}. We redo the linear
 stability analysis(LSA) with a physically well motivated angular emission spectrum, where the 
 $\nu_e$, which decouple later than the $\bar{\nu}_e,~\nu_x$ , have a larger flux and
 wider angular distribution than the latter. We study possible instabilities
 for neutrino flavor evolution in
 space as well as time, and for the first time, include backward going modes also. 
 Finally, we verify our LSA results with numerical results from the fully non-linear 
 evolution. 
 \section{Set-up of the problem}
 \label{sec:2}
Neutrino flavor evolution in a dense media is explained with space-time dependent Wigner functions
 $\varrho_{{\bf p}, {\bf x},t}$ with momentum ${\bf p}$ at position ${\bf x}$ and time $t$. The equation
 of motion (EoM) is 
 \begin{equation}
\partial_t \varrho_{{\bf p}, {\bf x},t} + {\bf v}_{\bf p} \cdot \nabla_{\bf x} \varrho_{{\bf p}, {\bf x},t} 
= - i [\Omega_{{\bf p}, {\bf x},t}, \varrho_{{\bf p}, {\bf x},t}]\nonumber
\,\ ,
\label{eq:eom}
\end{equation}
 Here we have neglected external forces acting on the system as well as collisions. The Hamiltonian
 matrix is $\Omega_{{\bf p}}= \Omega_{{\rm vac}} + \Omega_{\rm MSW} + \Omega_{\nu\nu} \,\ ,$
where $ \Omega_{{\rm vac}} = \textrm{diag}(-\omega/2, +\omega/2)\,$ is the vacuum term with $ \w=\Delta m^2/2E $;
 $\Omega_{\rm MSW} =  \lambda\,\ \textrm{diag} (1,0) \,$ is the matter term with  $\lambda =\sqrt{2} G_F n_e$, where $n_e$ is the electron density;  and
 $\Omega_{\nu\nu}= \sqrt{2} G_F \int \frac{d^3 {\bf q}}{(2 \pi)^3} ({\varrho_{\bf q}} - {\bar\varrho_{\bf q}}) (1 -{\bf v}_{\bf p}\cdot {\bf v}_{\bf q})$ is
 the multi-angle neutrino-neutrino interaction term. Since the total number of neutrinos is always conserved, we can write 
 \begin{figure}[!h]
\begin{centering}
\includegraphics[width=0.3\textwidth]{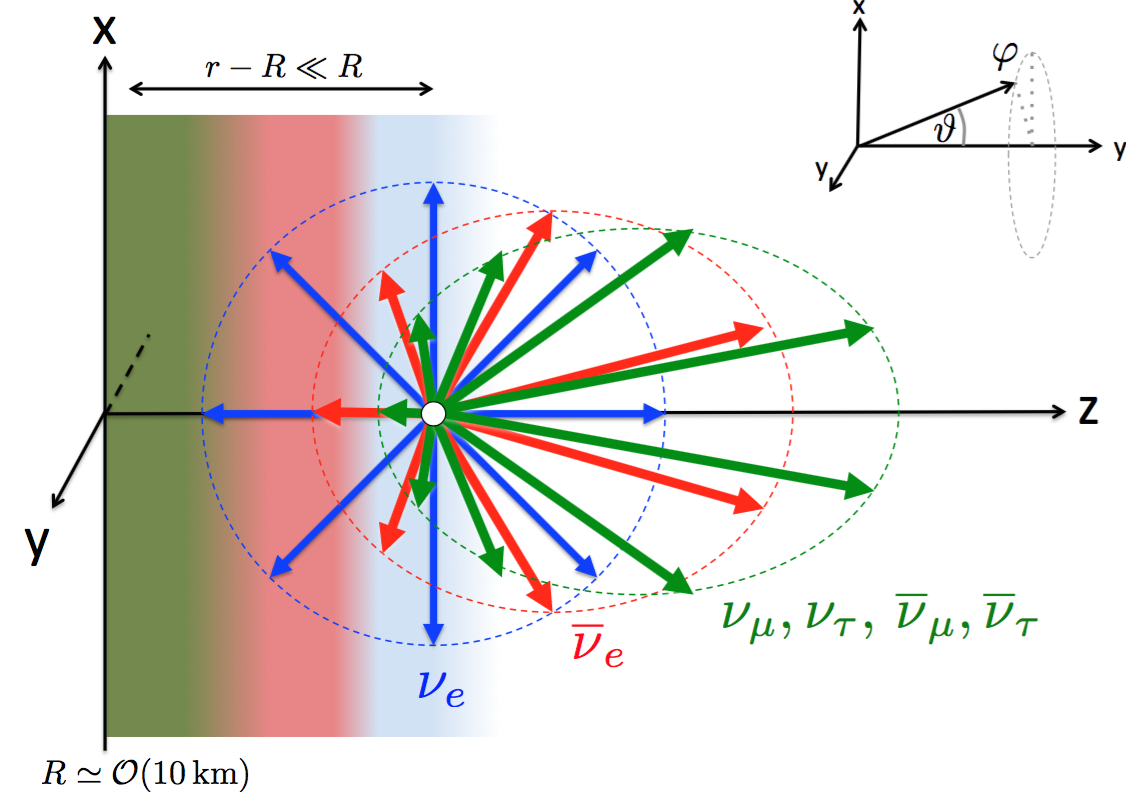}\hspace{1.0cm}\includegraphics[width=0.24\textwidth]{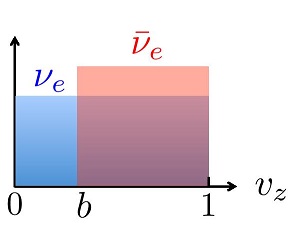}\hspace{1cm}\includegraphics[width=0.34\textwidth]{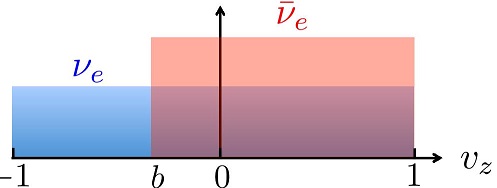}
\end{centering}
\caption{Left panel shows schematic polar plots of the angular distributions of the $\nu_e$ (blue), $\bar{\nu}_e$ (red),
and $\nu_x$ (green) emission fluxes. Middle panel shows a spectrum with no ingoing $\nu_e$ or $\bar\nu_e$ , while the right panel shows ingoing $\nu_e$ and $\bar\nu_e$ .}
\label{fig:2}
\end{figure}  
 \begin{equation}
\varrho_{\omega,v_z,\varphi}=\frac{1}{2}{\rm Tr}(\varrho_{\omega,v_z,\varphi})\,{\mathbb I}+\Phi_\nu\frac{g_{\omega,v_z,\varphi}}{2}\begin{pmatrix}
                                                \,s_{\omega,v_z,\varphi} && S_{\omega,v_z,\varphi}\\
                                                S_{\omega,v_z,\varphi}^* && -s_{\omega,v_z,\varphi}
                                               \end{pmatrix}\,,\nonumber
\end{equation}
and drop the trace term. Here $S_{\omega,v_z,\varphi}\ll 1$ and $s_{\omega,v_z,\varphi}^2+|S_{\omega,v_z,\varphi}|^2=1$. Also $\Phi_\nu$ is the normalization of the differential spectrum $g_{\omega,v_z,\varphi}$, chosen
accordingly. As neutrinos are produced as
flavor eigenstates, no oscillation occurs as long as $S_{\omega,v_z,\varphi}=0$. We linearize the equations in this small parameter $S_{\omega,v_z,\varphi}$
to get an eigenvalue equation \cite{Banerjee:2011fj}
\begin{eqnarray}
  i(\partial_t+v_z\partial_z+\vec{v}_T\cdot\partial_{T})S_{\w,v_z,\varphi}&=&\left[\w+\lam+\mu\int d\Gamma'\left(1-v_zv_z'-\vec{v}_T.\vec{v}_T'\right)g_{\w',v_z',\varphi'} \right]S_{\w,v_z,\varphi} \nonumber\\
                                               &&- \mu\int d\Gamma'\left(1-v_zv_z'-\vec{v}_T.\vec{v}_T'\right)g_{\w',v_z',\varphi'}\,S_{\w',v_z',\varphi'}\,, \nonumber
\label{eq:stabeom}
\end{eqnarray}

where $\vec{v}_T$ is the velocity vector of the neutrino projected on the $x$-$y$--plane. The important quantity here is the difference in the 
diiferential spectrum, given by $g_{\omega,v_z,\varphi}\propto d\phi_{\nu_e}/d\Gamma-d\phi_{\nu_x}/d\Gamma$
for neutrinos and $\propto d\phi_{\nu_x}/d\Gamma-d\phi_{\bar\nu_e}/d\Gamma$ for antineutrinos.
 Keeping in mind that the angular spectrum of emission should be different for neutrinos and antineutrinos, we choose the following schematic spectrum
%\begin{equation}
$g_{\omega,v_z,\varphi}=\frac{1}{2\pi}\left[(1+a)\delta(\omega)\Theta(v_z)\Theta(1-v_z)-\delta(\omega)\frac{1}{(1-b)}\Theta(v_z-b)\Theta(1-v_z)\right]\,$
%\label{eq:spectrum}\nonumber
%\end{equation}
as shown in Fig.\,\ref{fig:2} (middle panel). Here $a$ denotes the neutrino-antineutrino asymmetry whereas $b$ controls the difference in zenith angle distribution.
Such a ``non-trivial'' angular emission spectrum seems to be crucial for fast conversion.

To solve the eigenvalue equation, we take $S= Qe^{-i(\Omega_t t+\Omega_z z)}$, where $\Omega_{(t,z)}=\gamma_{(t,z)}+i\kappa_{(t,z)}$ can take complex values. A non-zero positive $\kappa_{(t,z)}$ causes
an exponential growth in $S$, thereby signalling an instability. It is important to mention that since we are looking for fast conversions, we can integrate out $\w$ from the spectrum and effectively set 
$\w/\mu \rightarrow 0$.
\section{Results}
 \label{sec:3}
\begin{figure}
\begin{centering}
\includegraphics[width=0.32\textwidth]{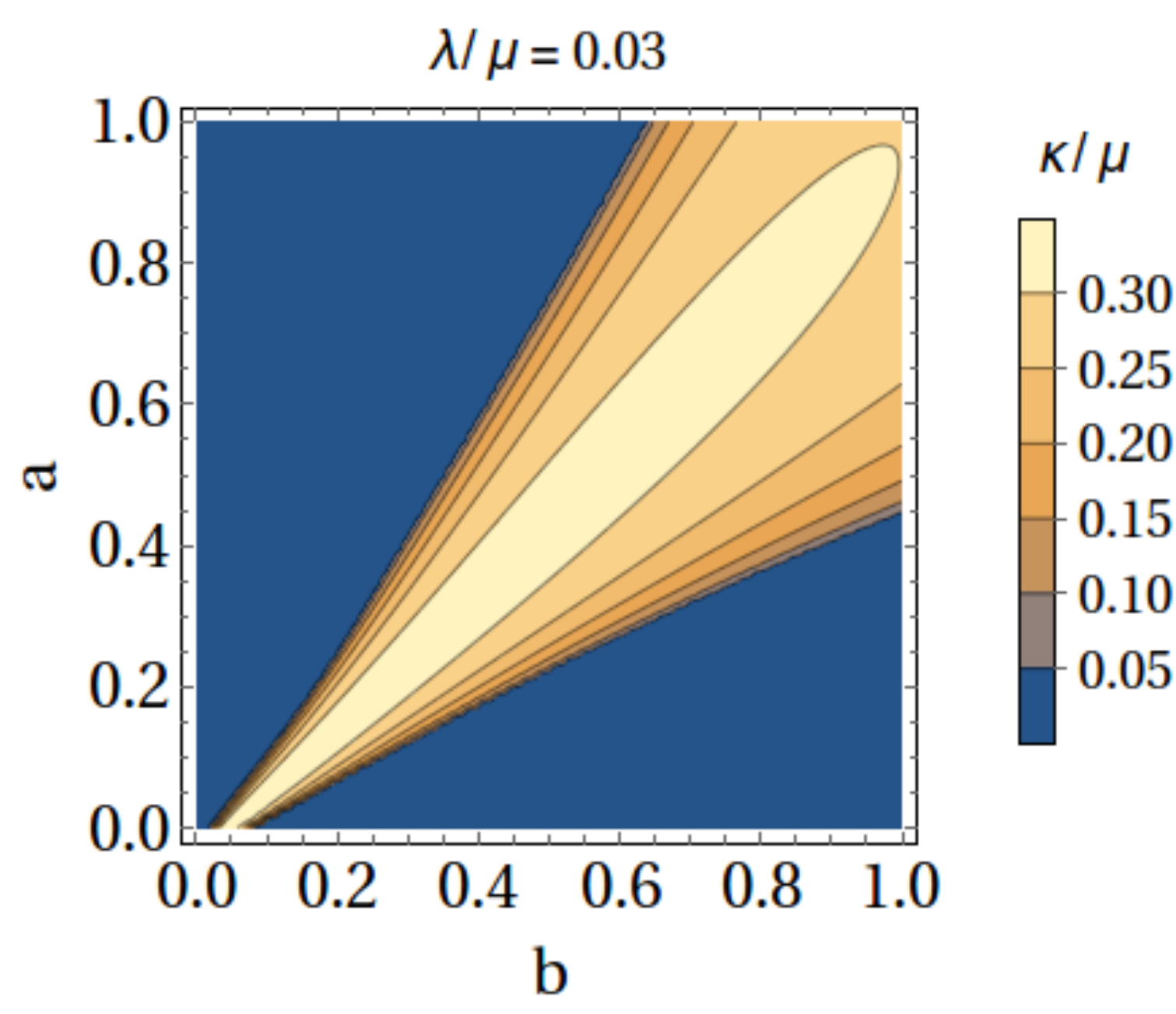}~\includegraphics[width=0.32\textwidth]{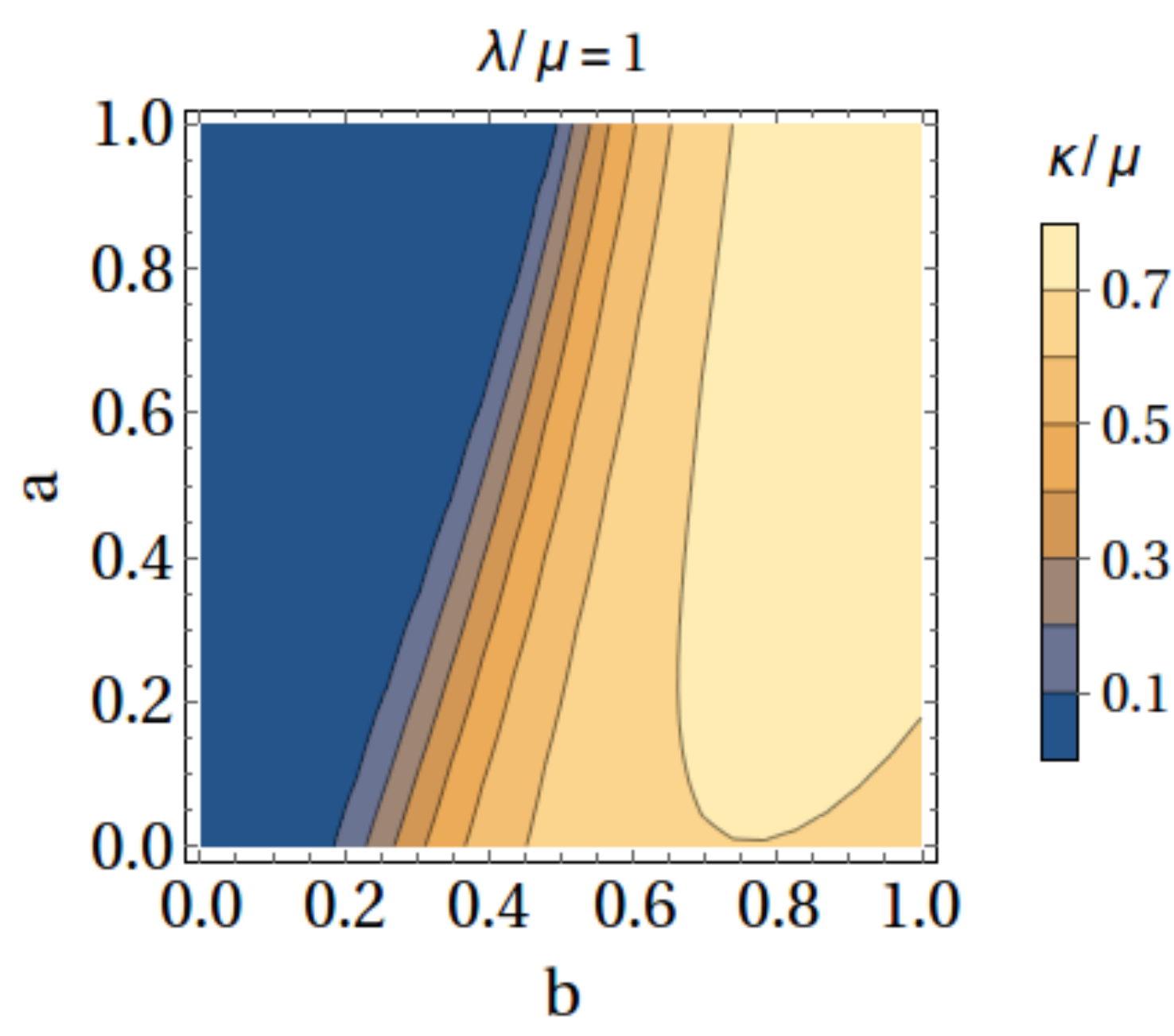}~\includegraphics[width=0.33\textwidth]{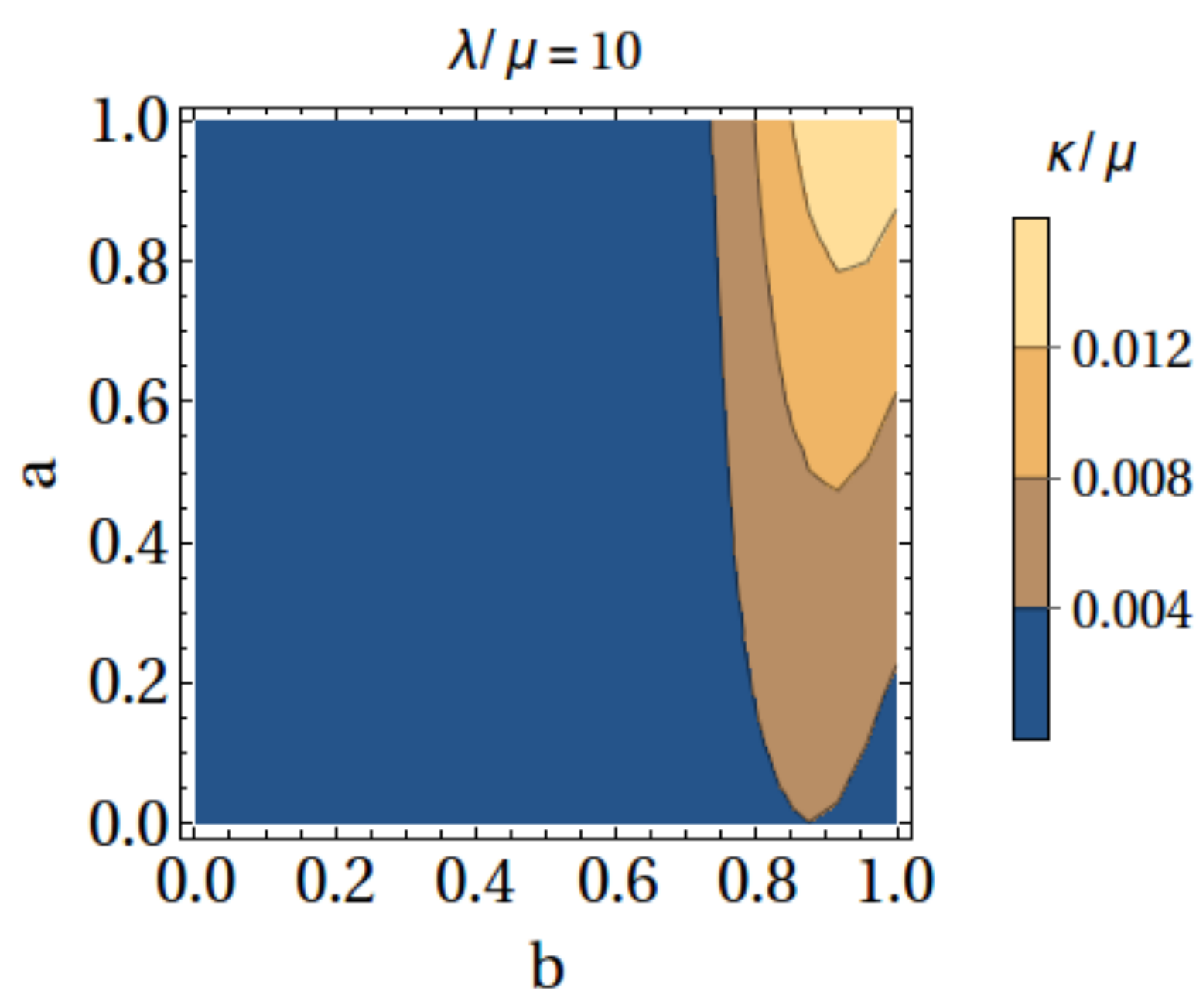}
\end{centering}
\caption{Instability growth rates for evolution in space for three different values of $\lambda/\mu=$ 0.03, 1, and 10, respectively from left to right.}
\label{fig:3}
\end{figure}

 Armed with this formalism, we look for instabilities for evolution in time $(\Omega_z\rightarrow0)$ and in space $(\Omega_t\rightarrow0)$.  We show a contour plot 
 $\kappa_z$ for different values of $a$ and $b$ in Fig.\,\ref{fig:3}. A common feature of all these
 plots is that no fast conversion takes place if $b=0$, thereby indicating that a non-trivial spectrum might be necessary for fast conversions. Also, matter suppresses these growths. 
\begin{figure}[t]
\begin{centering}
\includegraphics[width=0.35\textwidth]{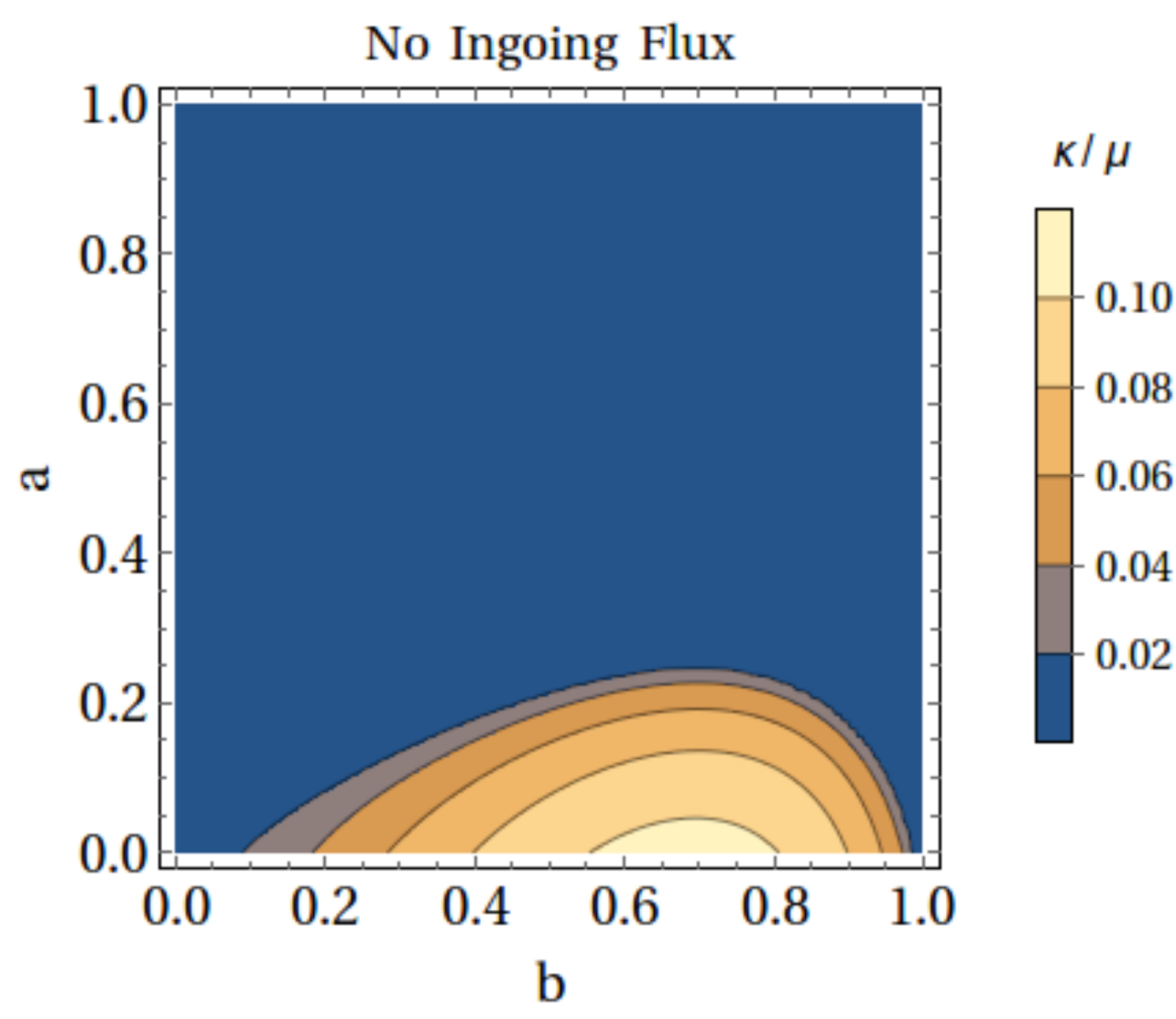}\hspace{2.0cm}\includegraphics[width=0.35\textwidth]{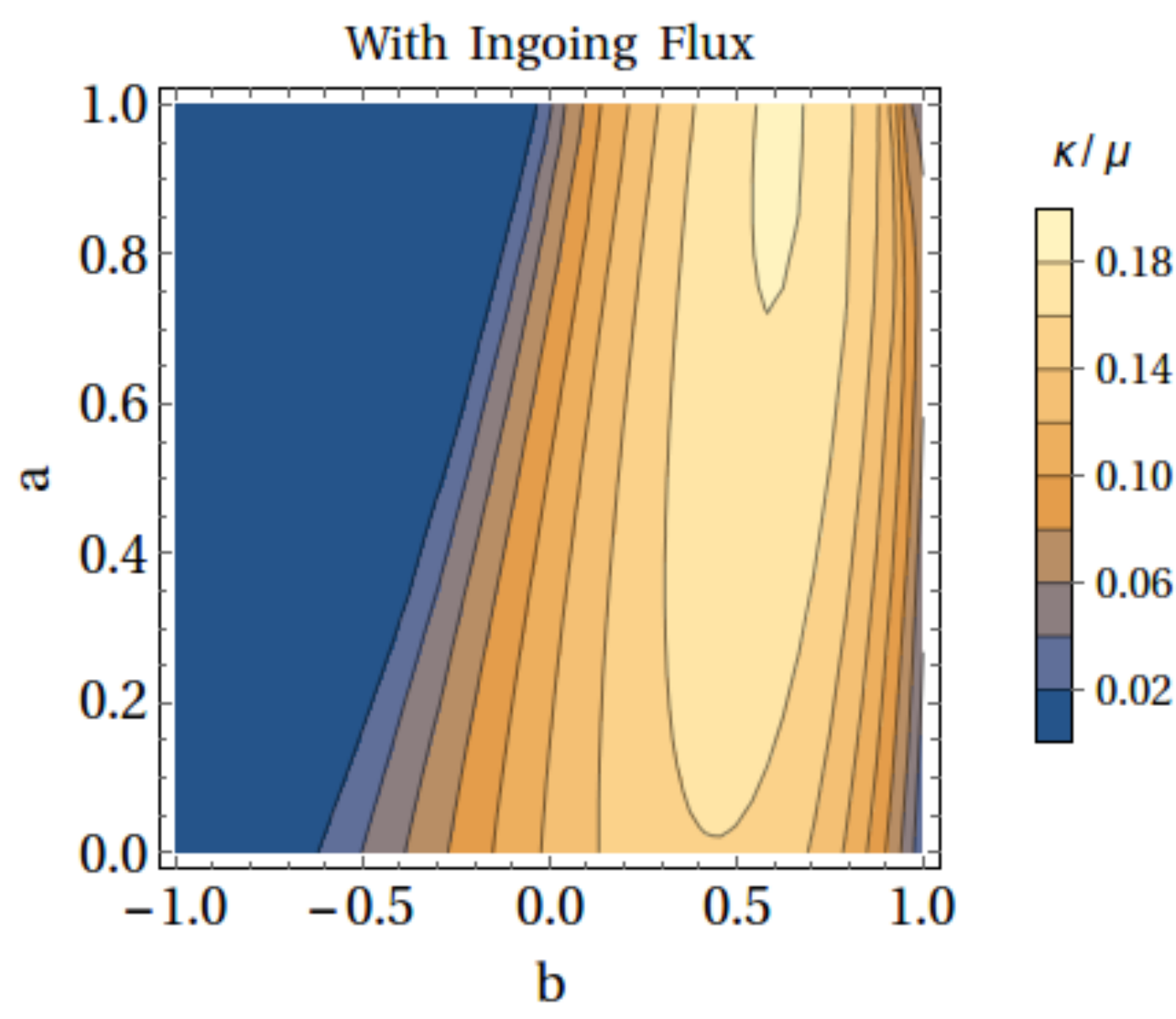}
\end{centering}
\caption{Instability growth rates for evolution in time. Left: without inward going modes. Right: with inward going modes.}
\label{fig:4}
\end{figure}
\begin{figure}[!t]
\begin{centering}
\includegraphics[width=0.4\textwidth]{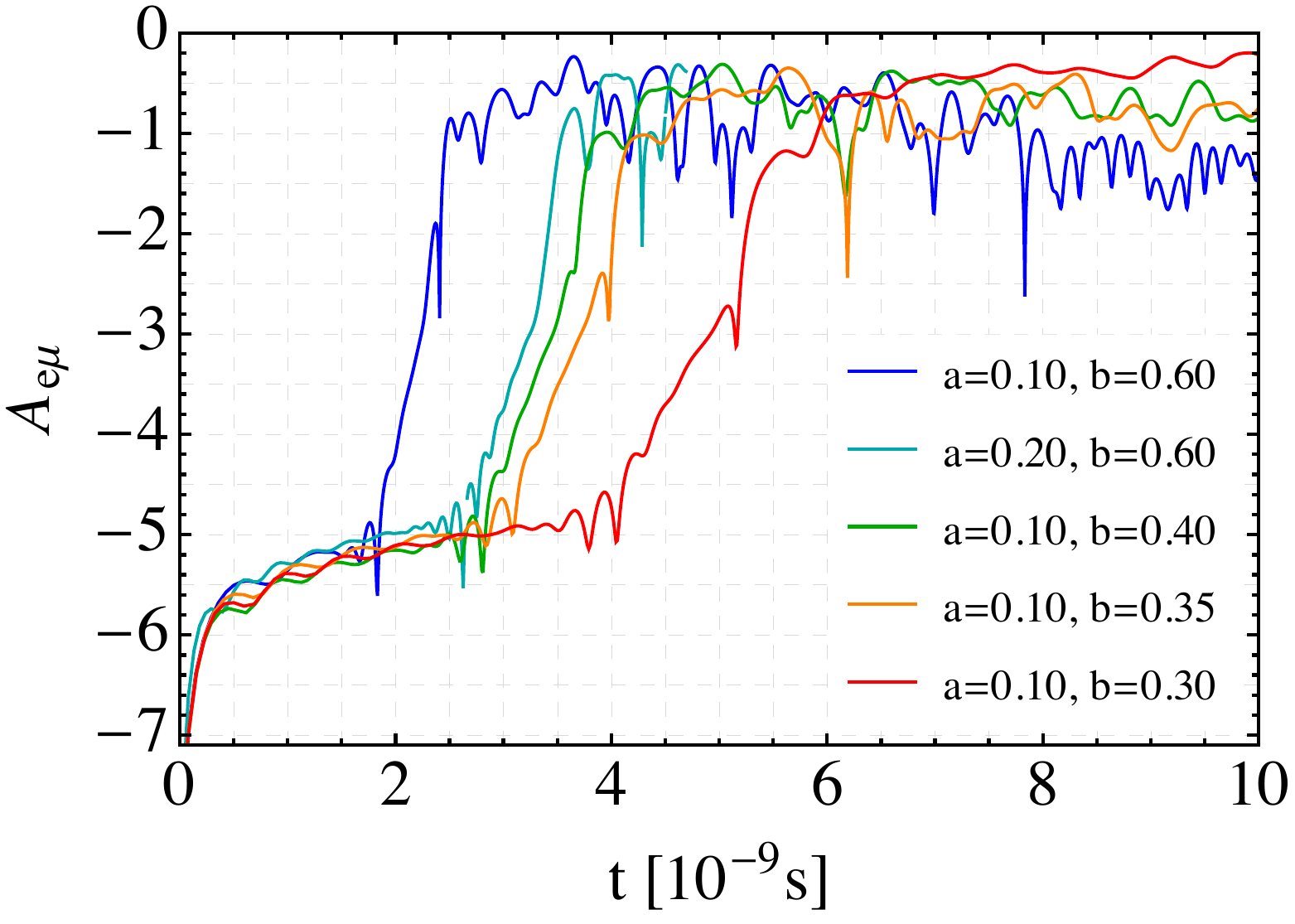}\hspace{2.0cm}\includegraphics[width=0.4\textwidth]{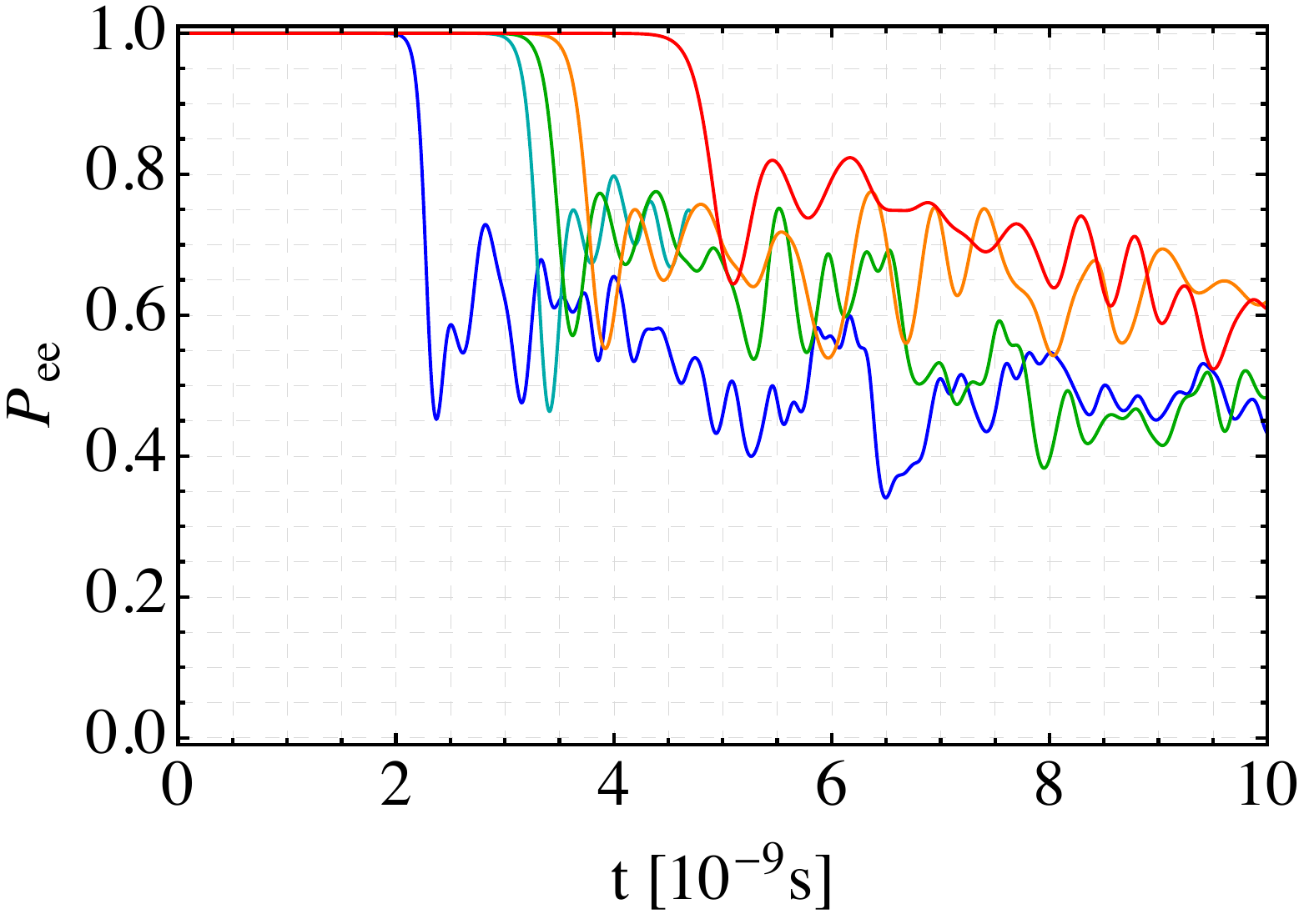}
\end{centering}
\caption{Instability growth for evolution in time. Left panel shows the quantity $A_{e\mu}={\log}_{10}|S|$ giving the extent of flavor conversion. 
Right panel shows the electron neutrino survival probabilities~$P_{ee}$ for same values of $a$ and $b$.}
\label{fig:5}
\end{figure}
Similarly, in Fig.\,\ref{fig:4}, we plot the growth rates for $\kappa_t$. We also show similar instabilities for a  spectrum including backward travelling modes as
$g_{\omega,v_z,\varphi}=\frac{1}{2\pi}\left[\frac{1+a}{2}\delta(\omega)\Theta(1+v_z)\Theta(1-v_z)-\delta(\omega)\frac{1}{(1-b)}\Theta(v_z-b)\Theta(1-v_z)\right]$
as shown in Fig.\,\ref{fig:2} (right panel). We note that inclusion of backward going modes increases fast conversions. Interestingly, matter suppression does not occur for time evolution, as suggested in \cite{Dasgupta:2015iia}.
 To verify our results from LSA, we have also solved the fully non-linear EoMs for specific values of $a$ and $b$.
The results shown in Fig.\,\ref{fig:5} indicate that indeed complete flavor averaging takes place within a few nanoseconds. One is also led to speculate whether it is important to have a crossing in the angular spectrum for development 
of fast conversions, as is evident from all the above cases. Further details are worked out in \cite{Dasgupta:2016dbv}.

 \section{Conclusion}
 \label{sec:4}
Thus we notice that for a non-trivial distribution of angular spectrum, we can indeed get rapid flavor turn-overs in time, even if in space the growth is suppressed by matter effects.
Also, backward travelling modes near the neutrinosphere aid in fast conversion.  If these fast conversions indeed take place so near the SN core, they may have important implications
for supernova explosion mechanism and nucleosynthesis. 



\begin{thebibliography}{6}
%
%\cite{Wolfenstein:1977ue}
\bibitem{Wolfenstein:1977ue} 
  L.~Wolfenstein,
  %``Neutrino Oscillations in Matter,''
  Phys.\ Rev.\ D {\bf 17}, 2369 (1978).
  doi:10.1103/PhysRevD.17.2369
  %%CITATION = doi:10.1103/PhysRevD.17.2369;%%
  
  %\cite{Mikheev:1986gs}
\bibitem{Mikheev:1986gs} 
  S.~P.~Mikheev and A.~Y.~Smirnov,
    %``Resonance Amplification of Oscillations in Matter and Spectroscopy of Solar Neutrinos,''
  Sov.\ J.\ Nucl.\ Phys.\  {\bf 42}, 913 (1985)
  [Yad.\ Fiz.\  {\bf 42}, 1441 (1985)].
  %%CITATION = SJNCA,42,913;%%
    %\cite{Duan:2006an}
\bibitem{Duan:2006an} 
  H.~Duan, G.~M.~Fuller, J.~Carlson and Y.~Z.~Qian,
  %``Simulation of Coherent Non-Linear Neutrino Flavor Transformation in the Supernova Environment. 1. Correlated Neutrino Trajectories,''
  Phys.\ Rev.\ D {\bf 74}, 105014 (2006)
  doi:10.1103/PhysRevD.74.105014
  [astro-ph/0606616].
  %%CITATION = doi:10.1103/PhysRevD.74.105014;%%
  
    %\cite{Hannestad:2006nj}
\bibitem{Hannestad:2006nj} 
  S.~Hannestad, G.~G.~Raffelt, G.~Sigl and Y.~Y.~Y.~Wong,
 % ``Self-induced conversion in dense neutrino gases: Pendulum in flavour space,''
  Phys.\ Rev.\ D {\bf 74}, 105010 (2006)
  Erratum: [Phys.\ Rev.\ D {\bf 76}, 029901 (2007)]
  doi:10.1103/PhysRevD.74.105010, 10.1103/PhysRevD.76.029901
  [astro-ph/0608695].
  %%CITATION = doi:10.1103/PhysRevD.74.105010, 10.1103/PhysRevD.76.029901;%%
  %\cite{Sawyer:2015dsa}
\bibitem{Sawyer:2015dsa} 
  R.~F.~Sawyer,
  %``Neutrino cloud instabilities just above the neutrino sphere of a supernova,''
  Phys.\ Rev.\ Lett.\  {\bf 116}, no. 8, 081101 (2016)
  doi:10.1103/PhysRevLett.116.081101
  [arXiv:1509.03323 [astro-ph.HE]].
  %%CITATION = doi:10.1103/PhysRevLett.116.081101;%%

%\cite{Chakraborty:2016lct}
\bibitem{Chakraborty:2016lct} 
  S.~Chakraborty, R.~S.~Hansen, I.~Izaguirre and G.~Raffelt,
 % ``Self-induced neutrino flavor conversion without flavor mixing,''
  JCAP {\bf 1603}, no. 03, 042 (2016)
  doi:10.1088/1475-7516/2016/03/042
  [arXiv:1602.00698 [hep-ph]].
  %%CITATION = doi:10.1088/1475-7516/2016/03/042;%%
  
  %\cite{Banerjee:2011fj}
\bibitem{Banerjee:2011fj}
  A.~Banerjee, A.~Dighe and G.~Raffelt,
  %``Linearized flavor-stability analysis of dense neutrino streams,''
  Phys.\ Rev.\ D {\bf 84} (2011) 053013
  doi:10.1103/PhysRevD.84.053013
  [arXiv:1107.2308 [hep-ph]].
  %%CITATION = doi:10.1103/PhysRevD.84.053013;%%
  %62 citations counted in INSPIRE as of 21 Feb 2017
  
  
 \bibitem{Dasgupta:2016dbv}
  B.~Dasgupta, A.~Mirizzi and M.~Sen,
  %``Fast neutrino flavor conversions near the supernova core with realistic flavor-dependent angular distributions,''
  arXiv:1609.00528 [hep-ph].
  
  %\cite{Dasgupta:2015iia}
\bibitem{Dasgupta:2015iia}
  B.~Dasgupta and A.~Mirizzi,
  %``Temporal Instability Enables Neutrino Flavor Conversions Deep Inside Supernovae,''
  Phys.\ Rev.\ D {\bf 92} (2015) no.12,  125030
  doi:10.1103/PhysRevD.92.125030
  [arXiv:1509.03171 [hep-ph]].

\end{thebibliography}
\end{document}